\documentclass[aps, prl, amsmath, amssymb, preprintnumbers, twocolumn, showpacs, superscriptaddress]{revtex4-1}
\usepackage{amssymb, amsthm}
\usepackage{color}
\usepackage{graphicx}
\usepackage{subfigure}
\usepackage{amssymb}
\usepackage{ulem}
\usepackage{comment}
\usepackage{hyperref}
\usepackage{esint}
\usepackage{amsmath}
\usepackage{ulem} 

\newcommand{\ket}[1]{| #1 \rangle}
\newcommand{\bra}[1]{\langle #1 |}

\definecolor{grey}{rgb}{.6,.6,.6}

\definecolor{orange}{rgb}{1,0.5,0}

\newcommand{\mytitle}{Topological phases of parafermions: a model with exactly-solvable ground states}

\begin{document}

\title{\mytitle}

\author{Fernando Iemini}
\affiliation{ICTP, Strada Costiera 11, I-34151 Trieste, Italy}
\affiliation{NEST, Scuola Normale Superiore and Istituto Nanoscienze-CNR, I-56126 Pisa, Italy}

\author{Christophe Mora}
\affiliation{Laboratoire  Pierre  Aigrain,
\'Ecole  Normale  Sup\'erieure / PSL  Research  University,
CNRS,  Universit\'e  Pierre  et  Marie  Curie-Sorbonne  Universit\'es,
Universit\'e  Paris  Diderot-Sorbonne  Paris  Cit\'e,  24  rue  Lhomond,  75231  Paris  Cedex  05,  France}

\author{Leonardo Mazza}
\affiliation{D\'epartement de Physique, \'Ecole Normale Sup\'erieure / PSL Research University, CNRS, 24 rue Lhomond, F-75005 Paris, France}

\begin{abstract}
 Parafermions are emergent excitations that generalize Majorana fermions and can also realize topological order.
In this paper we present a non-trivial and quasi-exactly-solvable model for a chain of parafermions in a  topological phase. 
We compute and characterize the ground-state wavefunctions, which are matrix-product states and have a particularly elegant interpretation in terms of Fock parafermions, reflecting the factorized nature of the ground states.
Using these wavefunctions, we demonstrate analytically several signatures of topological order.
Our study provides a starting point for the non-approximate study of topological one-dimensional parafermionic chains with spatial-inversion and time-reversal symmetry in the absence of strong edge modes.
\end{abstract}


\maketitle

\paragraph{Introduction.}

The study of topological order (TO) is currently one of the most active research fields in condensed-matter physics. 
From the AKLT model~\cite{Affleck_1987} to the Laughlin wavefunction~\cite{Laughlin_1983}, from the Kitaev chain~\cite{Kitaev_2001} to the Toric code~\cite{Kitaev_2003}, this study has always benefited from the development of exactly-solvable models and of paradigmatic wavefunctions, whose detailed analysis permits the formation of a clear physical intuition, to be used in the understanding of complex experimental setups.

In this letter we focus on \textit{parafermions}, a generalization of Majorana fermions ~\cite{Alicea_2016}. 
After the experimental clarification that two zero-energy Majorana modes can be localized at the edges of a one-dimensional fermionic wire~\cite{Mourik_2012, Albrecht_2016}, the possibility of localizing parafermionic modes, and letting them interact, is currently under deep investigation. 
These excitations cannot appear in strictly one-dimensional spinless fermionic systems~\cite{Turner_2011, Fidkowski_2011}, but may emerge at the edge of a two-dimensional fractional topological insulator coupled to alternating ferromagnetic and superconducting materials~\cite{Alicea_2016, Barkeshli_2012, Lindner_2012, Cheng_2012, Clarke_2013, Vaezi_2013, Santos_2016}, as well as in other nanostructures or models~\cite{You_2012, Barkeshli_2013a, Barkeshli_2013b, Klinovaja_2014a, Klinovaja_2014b, Klinovaja_2014c, Barkeshli_2014a, Barkeshli_2014b, Barkeshli_2014c}. 

In these setups, one-dimensional chains of interacting parafermions arise, which, in certain circumstances, display  TO and edge $\mathbb Z_N$ parafermionic modes~\cite{Alicea_2016,  Burrello_2013, Motruk_2013, Bondesan_2013, Jermyn_2014, Milsted_2014, Stoudenmire_2015, Zhuang_2015, Sreejith_2016, Cobanera_2015,Fendley_JP2014}. 
Such edge modes are called strong when they commute with the Hamiltonian~\cite{Fendley_2012} and thereby generate a $N$-fold degeneracy in the entire spectrum, and weak when the commutation property and associated degeneracy are restricted to the ground state manifold. TO survives  weak  perturbations and hosts indistinguishably weak or strong modes~\cite{Alexandradinata_2015}. 
The importance of parafermionic zero-modes for topological quantum computation~\cite{Hutter_2016} motivates further investigations of these fractionalized systems.

In this letter we provide a non-trivial family of parafermionic models for which the properties of the ground states can be exactly characterized. These models are gapped, display TO, have spatial-inversion and time-reversal symmetries, and feature weak edge modes; 
they thus belong to the same symmetry class for which weak edge modes have been discussed so far with numerical and perturbative analytical methods~\cite{Alexandradinata_2015, Jermyn_2014, Zhuang_2015}, with the advantage of being easy to handle.
We analytically establish several key signatures of TO which can be easily extracted from the wavefunctions: (i) the presence of non-local edge-edge correlations, (ii) the indistinguishability of ground states by a symmetry-preserving local observable, (iii) the fact that only operators living at the edges are able to permute ground states, (iv) the $N$-fold degeneracy of the entanglement spectrum~\cite{Motruk_2013}. We also motivate the existence of weak edge modes.

The analysis rests on an intuitive ``particle-like'' picture of parafermions~\cite{Cobanera_2014, Cobanera_2014b}, that naturally leads to a formulation of the ground states in terms of matrix-product states (MPS)~\cite{PerezGarcia_2007}. 
Our model is thus a simple platform for the direct study of TO in parafermionic systems, which is particularly valuable given even the absence of a non-interacting and exactly-solvable limit (see however Ref.~\cite{Fendley_JP2014}).
For simplicity, we present our discussion in the case of $\mathbb Z_3$ parafermions, but the construction can be easily generalized to  $\mathbb Z_N$ parafermions. A similar study has been discussed in the fermionic ($\mathbb Z_2$) case~\cite{Katsura_2015}.

\paragraph{The model.}
We consider a one-dimensional chain with length $L$ of $\mathbb Z_3$ parafermions. Each lattice site $k$ is associated with two parafermionic operators, $\hat \gamma_{2k-1}$ and $\hat \gamma_{2k}$, which satisfy the following properties:
$ \hat \gamma_j^3 = 1$, $\hat \gamma_j^\dagger = \hat \gamma_j^2$; moreover, $ \hat \gamma_j \hat \gamma_l = \omega \, \hat \gamma_l \hat \gamma_j$ for $j<l$, where 
 $\omega = e^{2 \pi i /3}$.
We consider the following model: $\hat H = \hat H_0 + b \, \hat H_1 + b^2 \, \hat H_2$:
\begin{subequations}
\label{eq:Ham}
\begin{align}
 \hat H_0 =&  \sum_j \left[ 
 -f \, \omega^{*} \, \hat \gamma_{2j-1}^\dagger \hat \gamma_{2j} 
 -J 
 {\omega} \,
 \hat \gamma_{2j} \, \hat \gamma_{2j+1}^\dagger
 + \text{H.c.} \right]; \\
 \hat H_1 =& -J\sum_j  \left[ \hat A_{j}
 \hat \gamma^\dagger_{2j+1}+ \hat \gamma_{2j} \hat B^\dagger_{j+1} + \text{H.c.}  \right] ;\\
 \hat H_2 =& -J \sum_j \left[ \omega^*  \,
 \hat A_{j} \hat B_{j+1}^\dagger
 + \text{H.c.} \right];
\end{align}
\end{subequations}
and
$\hat A_{j} = ( \hat \gamma_{2j-1} +  \hat \gamma_{2j-1}^\dagger \hat \gamma_{2j}^\dagger )$,
$\hat B_{j} = ( \hat \gamma_{2j}+ \hat \gamma_{2j}^\dagger \hat \gamma_{2j-1}^\dagger )$.

For $b=0$, $\hat H$ reduces to the well-known parafermionic version of the three-state Potts quantum chain~\cite{Ostlund_1981, Huse_1981, Fateev_1982, Alcaraz_1987, Ortiz_2012, Mong_2014}.
For positive $f$ and $J$, such model has a topological phase transition at $f=J$ between a topological phase with zero boundary-modes ($f<J$) and a trivial phase ($f>J$). 
For $f=0$, the Hamiltonian is the sum of commuting and frustration-free terms, and displays TO.

\paragraph{Quasi-exactly-solvable line}

\begin{figure}[t]
 \includegraphics[width=\columnwidth]{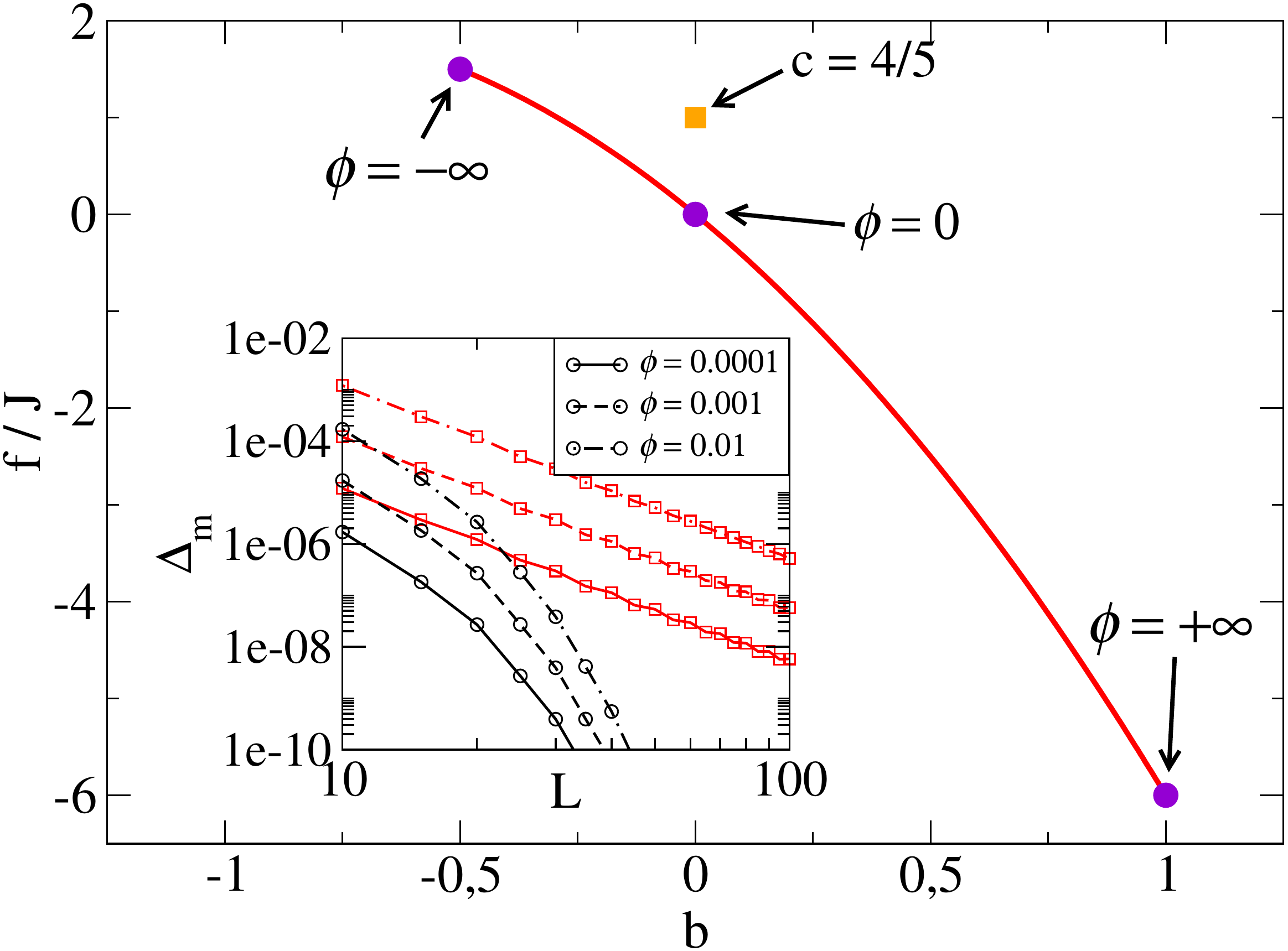}
 \caption{Phase space of the model~\eqref{eq:Ham}. 
 The ground state is exactly solvable along the red line, which is parametrized $\phi$ according to Eq.~\eqref{eq:exact:values}. The points $\phi = \pm \infty$ and $\phi = 0$ are highlighted. For better reference, the well-studied critical point $f = J$, $b = 0$ with central charge $c = 4/5$ is also highlighted.
 Inset: Perturbative analysis of the size-scaling of the degeneracies $\Delta_m$ of the first three excited states ($m=1$, black circles) and of a higher excited triplet ($m=4$, red squares), exhibiting respectively exponential and polynomial energy splitting. The polynomial scaling demonstrates the absence of strong edge modes. Three values of $\phi$ are considered: $\phi = 10^{-4}$ (solid line), $\phi = 10^{-3}$ (dashed line), and $\phi = 10^{-2}$ (dashed-dotted line).}
 \label{Fig:Phase}
\end{figure}

The Hamiltonian~\eqref{eq:Ham} has a quasi-exactly-solvable line (where only the ground state but not the excited states can be exactly computed) parametrized by $\phi \in \mathbb R$:
\begin{equation}
 \frac fJ = -6 \frac{1 -e^{-2\phi}}{(1+2e^{- \phi})^2}; \qquad b = \frac{1-e^{-\phi}}{1+2e^{- \phi}} ; 
 \label{eq:exact:values}
\end{equation}
which is plotted in Fig.~\eqref{Fig:Phase}. 
We consider open boundary conditions; the properties of the ground states are exactly computable once the boundary term is introduced:
\begin{equation}\label{eq:boundary}
 \hat H_{\rm B} = + \frac f2 \left[\omega^{*} \hat \gamma_1^\dagger \hat \gamma_2+
 \omega^{*} \hat \gamma_{2L-1}^\dagger \hat \gamma_{2L}+ \text{H.c.} \right].
\end{equation}
This term does not change the thermodynamics  of the model and produces modifications which scale as $L^{-1}$, which are negligible in the thermodynamic limit.

We begin by considering the point $\phi = 0$. Here, the Hamiltonian can be rewritten in the following expressive form:
$
 \hat H+\hat H_{\rm B}  = 
 -2 J (L-1) \hat {\mathbb I}
 + J \sum_{j=1}^{L-1} \hat \ell_{j}^\dagger \hat \ell_{j}$,
where $\hat \ell_j = \hat \gamma_{2j}^\dagger-{\omega} \, \hat \gamma_{2j+1}^\dagger$.
The first term, inessential, is proportional to the identity. The second part, instead, is non-negative, and its three ground states $\ket{g_{i,\phi=0}} $ ($i = 0,1,2$) 
are characterized by $\hat \ell_j \ket{g_{i, \phi = 0}} = 0$.

In order to visualize this result in solely parafermionic terms, we employ the ``Fock parafermions'' $\{ \hat C_j \}_{j = 1}^L$: 
$ \hat \gamma_{2j-1} = {\omega \, \big(} \hat C_j + \hat C_j^{\dagger2} {\big)} $ and 
$ \hat \gamma_{2j} = \hat C_j \omega^{\hat N_j}+ \hat C_j^{\dagger2}$, where $\hat N_j = \hat C_j^\dagger \hat C_j +\hat C_j^{\dagger2} \hat C_j^2 $ is the number operator~\cite{Cobanera_2014}.
The Fock-parafermion operators are a generalization of canonical Fermi operators and satisfy, among the others, the following commutation relations: $\hat C_j^3 = 0$, $\hat C_j \hat C_k = \omega \, \hat C_k \hat C_j$ ($j<k$). They are associated with a local Fock space where a number of Fock-parafermions between $0$ and $2$ can be accommodated, and are amenable to a simple picture of particle-like excitations. The Hilbert space of the whole chain is spanned by all Fock states $\ket{\{ n_j\}}$ where $n_j \in \{0,1,2\}$ is the number of parafermions at site $j$.

The three ground states read:
\begin{equation}
 \ket{g_{i, \phi = 0}} = \frac{1}{\sqrt{3^{L-1}}}
 \sum_{\{ n_j\} \text{ such that}\atop \sum_j \hspace{-0.05cm} n_j \equiv i \; (\text{mod} 3)} \ket{\{ n_j\}},
 \quad i = 0,1,2.
 \label{eq:phi0:GS}
\end{equation}
They are the equal-amplitude superposition of all Fock states with a number $N$ of Fock parafermions such that $N \equiv i \; (\text{mod} 3 )$.
These states are similar to the Rokhsar-Kivelson states proposed in resonant valence-bond liquids~\cite{Rokhsar_1988}. Such states are in fact ubiquitous in the study of topological phases of matter and they can also be   encountered in the two- and three-dimensional toric code~\cite{Kitaev_2003, Lin_2015}, in the AKLT model~\cite{Affleck_1987} or in the study of topological Majorana zero-energy modes~\cite{Kitaev_2001, Lang_2015, Iemini_2015, Katsura_2015}. 
The proof of Eq.~\eqref{eq:phi0:GS} is obtained expanding $\hat \ell_j = \omega^{- \hat N_j} \hat C^\dagger_j - \hat C_{j+1}^\dagger+ \hat C^2_{j}-\hat C^2_{j+1}$, and explicitly inspecting that $\hat \ell_j \ket{g_{i,\phi = 0}}=0$.
Excited states are obtained by applying the operators $\hat \ell_j^\dagger$ to the  states $\ket{g_{i,\phi = 0}}$ and normalizing, which demonstrates the presence of a gap $3 J$~\cite{SuppMat}.

We now move to $\phi \neq 0$.
We claim that the ground states are given by
\begin{equation}
 \ket{g_{i, \phi}} = \frac{\hat Z_{- \phi} \ket{g_{i, \phi=0}}}{\sqrt{\bra{g_{i, \phi=0}} \hat Z_{- 2\phi} \ket{g_{i, \phi=0}}}},
 \label{eq:mod:GS}
\end{equation}
where $ \hat Z_{\phi} = e^{\phi \hat N /3} $ is a Hermitian, invertible, but non-unitary operator, $\hat N =\sum_j \hat N_j$ being the total number of parafermions in the chain. We prove our claim by constructing a parent Hamiltonian for the states $\ket{g_{i, \phi}}$ and then showing that it coincides with  $\hat H+ \hat H_{\rm B}$, as given by Eqs.~\eqref{eq:Ham} and~\eqref{eq:boundary}, apart from constant terms. We introduce a set of local operators $\hat L_{j,\phi}=\hat Z_{-\phi} \hat \ell_j  \hat Z_{\phi}$; one easily verifies that acting with the parent Hamiltonian 
\begin{equation}\label{hphi}
 \hat H_\phi = J \sum_{j = 1}^{L-1} \hat L_{j,\phi}^\dagger 
 \hat L_{j,\phi} \, 
\end{equation}
on the states $\ket{g_{i, \phi}}$ gives zero. The model is not fully solvable and the different terms in Eq.~\eqref{hphi} do not commute, except for $\phi=0$. Nevertheless $ \hat H_\phi $ is a strictly non-negative operator which completes the proof that $\ket{g_{i, \phi}}$ are ground states. 
More explicitely, the operators $\hat L_{j,\phi}$ take the form  $\hat L_{j,\phi} =\frac{e^{2 \phi/3}}{3} \left[ \hat {\mathcal W}_{j, \phi} \hat \gamma^\dagger_{2j}- \omega\hat {\mathcal W}_{j+1, \phi} \hat \gamma^\dagger_{2j+1}\right]$, with
$$ \hat {\mathcal W}_{j, \phi}  = (1+2 e^{-\phi})  + (1- e^{-\phi})\left[ {\omega} \, \hat \gamma_{2j-1}^\dagger \hat \gamma_{2j}+ \text{H.c.}\right], $$
such that $\hat H_\phi$ coincides with the starting Hamiltonian~\eqref{eq:Ham} and the parametrization~\eqref{eq:exact:values}.
The Hamiltonian remains time-reversal invariant (a detailed discussion is in~\cite{SuppMat}), as can be inferred by the fact that $\hat N_j$ satisfies such symmetry. Indeed, in usual parafermionic language $\hat N_j = 1 + [(\omega^*-\omega)\hat \gamma^\dagger_{2j-1} \hat \gamma_{2j} + \text{H.c.}]/3$, and since $\mathcal T$ is anti-unitary and maps $\mathcal T [\hat \gamma^\dagger_{2j-1} \hat \gamma_{2j}] = \hat \gamma^\dagger_{2j} \hat \gamma_{2j-1}$, the invariance follows.

\paragraph{Ground-state properties}
We now turn to the analytical characterization of the $\ket{g_{i,\phi}}$.
 In the Fock-parafermion representation, the ground states take a particularly simple form:
\begin{equation}
 \ket{g_{i,\phi}} = \frac{1}{\sqrt{\mathcal N_{L,\phi,i}}}
 \sum_{\{ n_j\} \text{ such that}\atop \sum_j \hspace{-0.05cm}n_j \equiv i \; (\text{mod} 3)} e^{- \phi (\sum_j n_j )/3} \ket{ \{ n_j\} },
 \label{eq:generic:GS}
\end{equation}
where the normalization constants $\mathcal N_{L,\phi,i}$ have an analytical expression. Comparing with the states in Eq.~\eqref{eq:phi0:GS}, the coefficients of the different Fock states now depend exponentially on the number of parafermions: the perturbation is effectively acting as a chemical potential which modifies the average number of particles. 

\begin{figure}[t]
\includegraphics[width=\columnwidth]{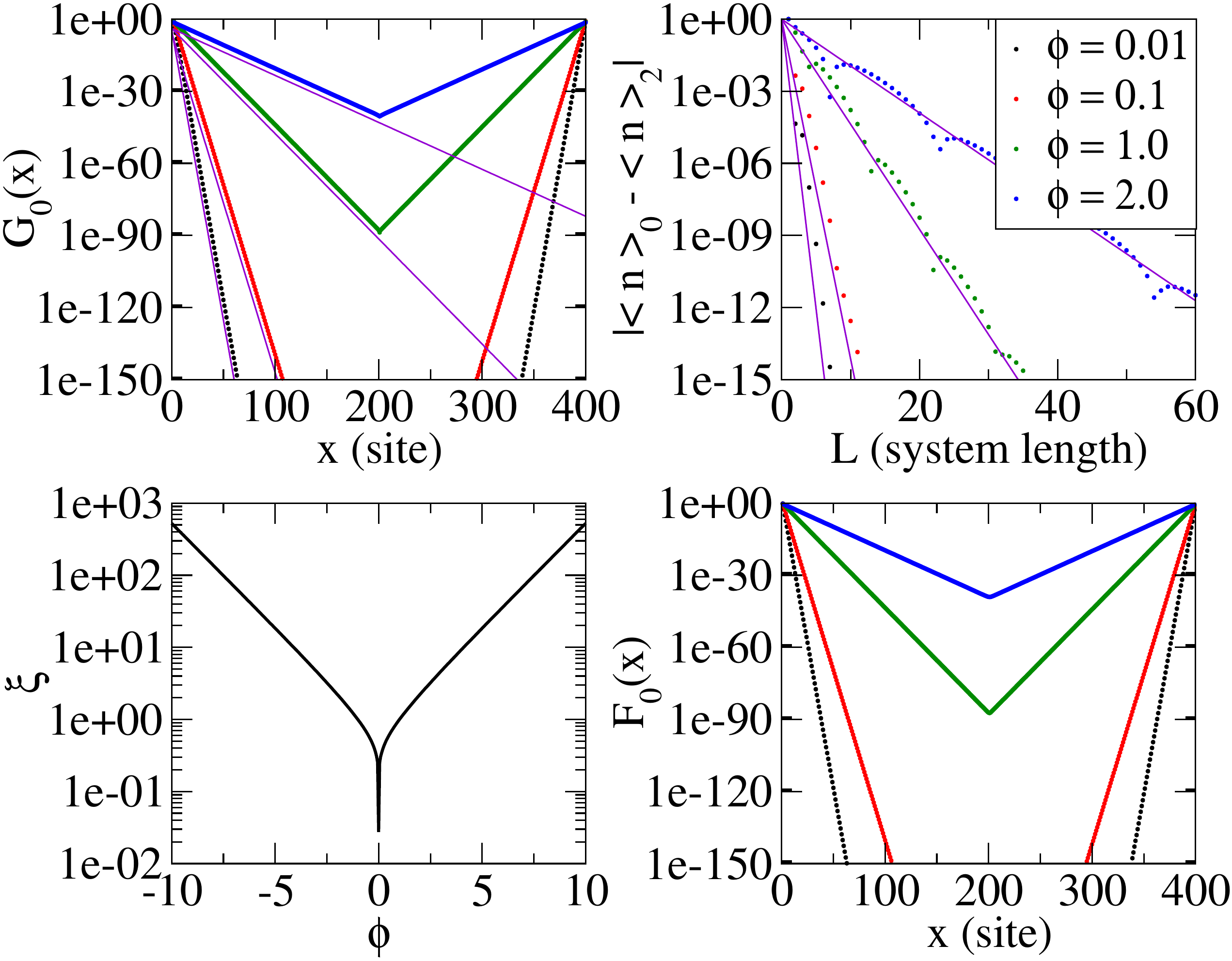}
\caption{Top left, correlation function $G_0(x)$, and Top right, $|\langle n \rangle_0 - \langle n \rangle_2|$ for several values of $\phi$. Violet lines represent the exponential scalings extracted from the analytical formulas.
Bottom left: correlation length $\xi$. Bottom right: we show one typical example of the $\mathbb Z_3$-breaking observable 
$F_{i}(x)$ for $i=0$.}
\label{fig:exact:analytical}
\end{figure}

Below, we take advantage of the relative simplicity of the ground states expressions~\eqref{eq:generic:GS} to compute analytically and exactly various correlation functions. We begin by determining the correlation length of the states $\ket{g_{i,\phi}}$ through a $\mathbb Z_3$-preserving correlation function, for instance $G_i(j,l) = \bra{g_{i,\phi}} \hat C^{\dagger 2}_j \hat C_l^2 \ket{g_{i,\phi}}$. 
The peculiar nature of the ground states makes it translationally invariant even for open boundary conditions,  $G_i (j,l) = G_i(j-l)$. 
As displayed in Fig.~\ref{fig:exact:analytical}, it exhibits an exponential decay $\sim \exp(- |j-l|/\xi)$ for $|j-l| < L/2$. $\xi$ is the correlation length:
\begin{equation}
 \xi^{-1} = \ln \left| \frac{1+e^{-2 \phi/3} +  e^{-4 \phi /3}}
 {1+ \omega e^{-2 \phi /3}+ \omega^* e^{- 4 \phi /3}} \right| ;
 \label{eq:corr:length}
\end{equation} 
and it is plotted in Fig.~\ref{fig:exact:analytical} as a function of $\phi$. It is zero at $\phi = 0$, corresponding to a renormalization group fixed point, and diverges in the limits $\phi \to \pm \infty$. 
Thus, no phase transition occurs along the solvable line, apart from the extremal values.

It is instructive to show that the correlation length can also be computed in ways that are directly related to the topological nature of the ground states.
We consider the expectation value of a $\mathbb Z_3$-preserving local operator, such as $\langle n \rangle_i = \bra{g_{i,\phi}} \hat N_j \ket{g_{i,\phi}}$ (for the states $\ket{g_{i,\phi}}$ there is no dependence on $j$).
In the thermodynamic limit, we analytically find:
\begin{equation}
\langle n \rangle_i \to n(\phi) =
 \frac{e^{-2 \phi/3} + 2 e^{-4 \phi /3}}
 {1+e^{-2 \phi /3}+e^{- 4 \phi /3}},
\end{equation}
independent of $i$ and $j$.
 At finite size $L$, we further obtain exponentially close values $\bra{g_{i,\phi}} \hat N_j \ket{g_{i,\phi}} = n(\phi) + c_i e^{-L/\xi}$, as expected for TO~\cite{Nussinov_2009}, with the  correlation length $\xi$ of Eq.~\eqref{eq:corr:length}.
  
Getting back to the correlation function $G_i(j-l)$  in Fig.~\ref{fig:exact:analytical},
we observe that the model displays non-local edge-edge correlations which survive in the thermodynamic limit.
The importance of the edges is also revealed by $\mathbb Z_3$-breaking observables: 
in Fig.~\ref{fig:exact:analytical}, we analytically compute and plot 
$F_{i}(j) = |\bra{g_{i,\phi}} \hat C^\dagger_j  \ket{g_{i-1 (\text{mod} 3),\phi}}|$ for $i=0$, measuring how $\hat C^\dagger_j $ maps groundstates with subsequent $\mathbb Z_3$ parities. The calculation reveals that it is non-zero only for $j$ close to the boundaries with exponential decays again characterized by $\xi$.
With this, we have so far encountered the first three signatures of  TO and fractionalized boundary modes mentioned in the introduction, points (i-iii).

\begin{figure}[t]
\includegraphics[width=\columnwidth]{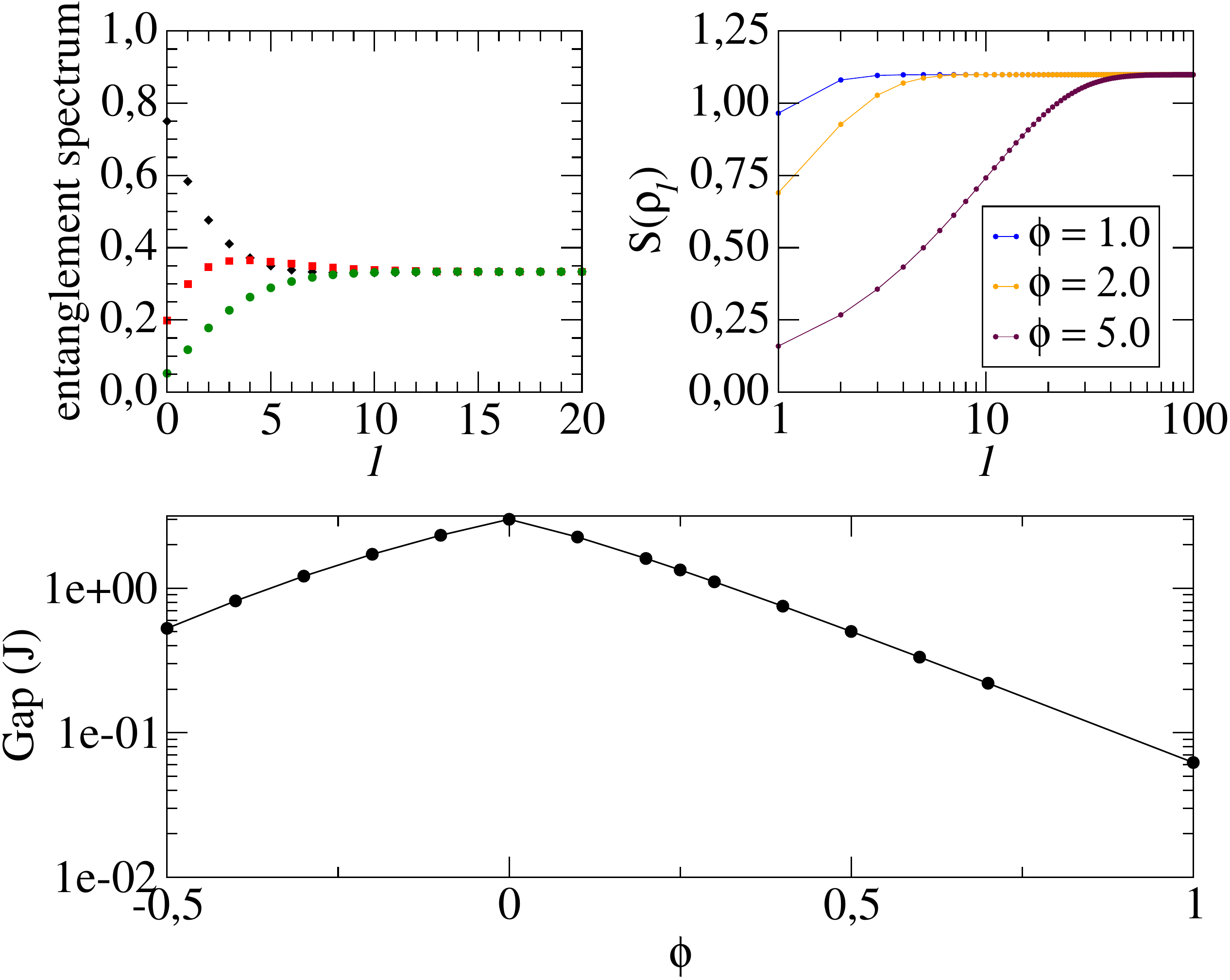}
\caption{Top left: entanglement spectrum as a function of $\ell$ for $\phi = 2$. 
Top right: von Neumann entropy $S(\hat \rho_\ell)$ as a function of $\ell$ for three values of $\phi$. At large   $\ell$, it saturates to a finite value corresponding to an area law. 
Bottom: DMRG calculation of the gap of the model obtained with length $L=168$; the maximal number of retained states is $m = 250$. }
\label{fig:exact:analytical:2}
\end{figure}

In order to confirm these findings, we consider the entanglement spectrum of the $\ket{g_{i,\phi}}$ states and prove its threefold degeneracy, see point (iv). In a bipartition of the system into a left part of length $\ell$ and a right part of length $L-\ell$, the ground state  assumes the form
\begin{equation}
\ket{g_{i,\phi}} = 
\sum_{p=0}^2
\sqrt{ \frac{ \mathcal N_{\ell,\phi,p} \, \mathcal N_{L-\ell,\phi,(i-p)\text{mod}3} } {\mathcal N_{L,\phi,i}}}
\ket{g^{(\ell)}_{\phi,p}}
\ket{g^{(L-\ell)}_{\phi,(i-p)\text{mod}3}}.
\label{eq:bipartition}
\end{equation}
The reduced density matrix $\hat \rho_\ell$ is obtained by tracing out all sites of the right part.
For $\ell \gg \xi$, the normalization constant $\mathcal N_{\ell,\phi,i}$ scales like $\sim (1/3)(1+e^{-2 \phi/3}+e^{-4\phi/3})^\ell +\mathcal O (e^{- \ell/\xi})$, and the dependence on $i$ only appears in the correction. Thus, for $\ell \gg \xi$ and $L-\ell \gg \xi$, the entanglement spectrum of the system is threefold degenerate, because for every $p= 0,1,2$ in Eq.~\eqref{eq:bipartition} the coefficient of the sum reduces to $\sqrt{1/3}$, and the three states $\ket{g^{(\ell)}_{\phi,p}}$ equally participate to the reduced density matrix $\hat \rho_\ell$.
In Fig.~\ref{fig:exact:analytical:2} we plot the typical behavior of the entanglement spectrum as a function of $\ell$, the position of bipartition: close to the boundary ($\ell \ll \xi$) it consists of three different values, away from it they all collapse to $1/3$. 
This expression also clarifies the gapped nature of the system through the area-law scaling of its von Neumann entropy $S(\hat \rho_\ell) = - \text{tr}[\hat \rho_\ell \ln \hat \rho_\ell]$, 
plotted in Fig.~\ref{fig:exact:analytical:2}. 
Explicit numerical calculations of the gap, obtained with the density-matrix renormalization group (DMRG)~\cite{Schollwoeck_2005}, reported in Fig.~\ref{fig:exact:analytical:2}, confirm this fact.

\paragraph{Nature of the ground states.}

This extended analytical analysis originates from the fact that the ground states are MPS. They can be expressed as 
$\ket{g_{i,\phi}} = \sum_{n_1, \ldots, n_L} v_L^T A^{[n_1]} \dots A^{[n_L]} v_{R,i} \ket{n_1, \ldots n_L}$ with the three matrices $A^{[j=0,1,2]} = (e^{-\frac{\phi}{3}} \hat \sigma)^j$, where
\begin{equation}
\hat \sigma = 
 \begin{pmatrix}
  1 & 0 & 0 \\
  0 & \omega & 0 \\
  0 & 0 & \omega^2 
 \end{pmatrix}.
\end{equation}
The parity $i$ of the ground state is encoded in the left and right vectors, with $v_L^T = (1,\omega^{i},\omega^{2 i})$ and $v_R^T = (1, 1, 1)$~\cite{PerezGarcia_2007}. 

A particularly clear interpretation of the data which we have so far displayed comes from the observation that the ground states $\ket{g_{i,\phi}}$ are linear superpositions of three product states, as we are going to show.
For $\phi = 0$, it can be explicitly verified that:
\begin{equation}
 \ket{g_{i,\phi=0}} = \frac{1}{\sqrt 3} \left( 
  \bigotimes_j \ket{\tilde 0_j} + \omega^i \bigotimes_j \ket{\tilde 1_j} + \omega^{2i } \bigotimes_j \ket{\tilde 2_j}\right); \nonumber
\end{equation}
where $\ket{\tilde i_j} = (\ket{n_j = 0}+ \omega^{\tilde{i}} \ket{n_j = 1}+ \omega^{2\tilde{i}} \ket{n_j = 2})/\sqrt 3$.
The operator $\hat Z_{\phi}$ acts as a product operator over the different sites, without creating entanglement or correlations. 
We thus observe that, applying $\hat {Z}_{-\phi}$ to the states $\ket{g_{i,\phi = 0}}$ according to the prescription in Eq.~\eqref{eq:mod:GS},
the states $\hat Z_{-\phi} \bigotimes_j \ket{\tilde i_j}$ retain a product nature. These states have zero correlation length and thus are fixed points of the renormalization group. 
This result can be considered an extension to 3-state clock models of known results for spin-1/2 systems about the existence of factorized ground states~\cite{Kurmann_1982, Mueller_1985} and it is intriguing to speculate that the peculiar properties of these models might extend to parafermionic chains~\cite{Giampaolo_2010}. 

\paragraph{Edge modes.} 
The parafermionic chain at $\phi=0$ is characterized by (strong) edge modes, simply given by the operators $\hat \chi_1 = \hat \gamma_1$ and  $\hat \chi_2 = \hat \gamma_{2 L}$, permuting cyclically the ground states. As $\phi$ departs from zero, the edge modes $\hat \chi_{1,2}$ are continuously deformed but remain local. The  calculation of $F_{i} (j)$, see Fig.~\ref{fig:exact:analytical}, already demonstrates that they keep a significant overlap with operators located close to the two ends of the chain. More generally, $\hat \chi_{1,2} = \hat {\mathcal{V}}_{\phi} \hat{\gamma}_{1,2 L} \hat{\mathcal{V}}_{\phi}^\dagger $ are obtained from exact quasi-adiabatic continuation~\cite{Alexandradinata_2015} of $\hat \gamma_{1,2 L}$ with the unitary transformation $\hat{\cal V}_\phi$ mapping the ground states manifolds at zero and non-zero $\phi$. Being unitary, $\hat{\cal V}_\phi$ preserves the parafermionic non-commutative algebra of $\hat \chi_{1,2}$ in the ground state, with $\bra{g_{i,\phi}} \hat{\chi}_1 \ket{g_{i+1,\phi}} =1$ and  $\bra{g_{i,\phi}} \hat{\chi}_{2} \ket{g_{i+1,\phi}} = \omega^{2+i}$ . It also preserves locality under the condition of quasi-adiabaticity~\cite{Alexandradinata_2015}. This program can be applied explicitely in perturbation with  $\phi \ll 1$, and yields the left edge mode
\begin{equation}\label{eq:bound}
\hat{\chi}_1 = \hat{\gamma}_1 +  \alpha ( \omega \gamma_3 - \gamma_2^\dagger \gamma_3^\dagger) + \alpha^* (\omega \gamma_1^\dagger  \gamma_3 \gamma_2 - \gamma_1^\dagger \gamma_3^\dagger )
\end{equation}
to leading order in $\alpha = f/J - \omega b$. Parafermionic operators $\hat \gamma_j$ are also expected to enter the expression of $\hat{\chi}_1$
to order $j/2$ in $f/J$, $b$, so they are exponentially suppressed with the site index $j$. Similar considerations apply to the right mode $\hat \chi_2$. Finally, the form of $F_{i} (j)$ strongly suggests that the edge states $\hat \chi_{1/2}$ decay with the correlation length $\xi$ at both ends of the chain.

The operators $\hat \gamma_1$ and  $\hat \gamma_{2 L}$ are strong edge modes for $\phi=0$ as they commute with the Hamiltonian. It can be checked from the perturbative expression~\eqref{eq:bound} that the commutation is lost at non-zero $\phi$, yielding weak edge modes. Following Ref.~\cite{Jermyn_2014}, the low-energy part of the spectrum can be addressed at small $\phi$ by projecting the full Hamiltonian~\eqref{eq:Ham} onto single domain wall excitations, thus reducing the numerical complexity. 
In Fig.~\ref{Fig:Phase} we show the results of this analysis, where we plot 
$ \Delta_m = \sqrt{ \sum_{q \neq q'} (e_{m,q} - e_{m,q'})^2  }$, and $e_{m,q}$ is the $m$-th excited state which has a $\mathbb Z_3$-parity with value $q$.
We find that the lowest triplets in the excitation spectrum ($m=1$) exhibit an exponential closing with the system size, in contrast with higher triplet excitations ($m=4$) where the closing is polynomial.
This last observation rules out the presence of strong edge modes.

\paragraph{Conclusions.}

In this article we have presented a model for a one-dimensional parafermionic chain which displays  TO and has quasi-exactly-solvable ground states. 
It beautifully exemplifies the physics proposed in Ref.~\cite{Alexandradinata_2015}, and
allows for the explicit characterization of TO in the absence of strong boundary modes,  providing an interesting starting point for developing a physical intuition of parafermionic systems, which is becoming particularly compelling in view of a forthcoming experimental realization.
This is achieved with a systematic interpretation of our results in terms of ``Fock parafermions'', a possibility which has not been fully explored yet. 
Understanding whether there are other parent Hamiltonians for the states $\ket{g_{i,\phi}}$ which are more physically relevant is an interesting perspective~\cite{Peschel_1986};
we also leave for the future further investigations concerning higher-dimensional lattices~\cite{Burrello_2013}, as well as the complete mapping of the phase diagram of Hamiltonian~\eqref{eq:Ham}.

\acknowledgments

{\it Acknowledgements.} We thank R.~Fazio for encouraging us to perform this study. We also thank C.~E.~Bardyn, M.~Burrello, M.~Dalmonte, A.~De~Luca, M.~Fagotti, G.~Ortiz, N.~Regnault, D.~Rossini, R.~Santachiara for interesting discussions and comments.
The numerical part of this work has been performed using the DMRG code released within the ``Powder with Power'' project (http://qti.sns.it/dmrg/home.html).
F.~I.~acknowledges  financial support  by
the  Brazilian  agencies  FAPEMIG,  CNPq,  and  INCT-IQ.
C.~M.~acknowledges support from Idex PSL Research University (ANR-10-IDEX-0001-02 PSL).
L.~M.~was supported
by LabEX ENS-ICFP: ANR-10-LABX-0010/ANR-10-
IDEX-0001-02 PSL*.

\clearpage
\newpage

\clearpage
\setcounter{equation}{0}%
\setcounter{figure}{0}%
\setcounter{table}{0}%
\renewcommand{\thetable}{S\arabic{table}}
\renewcommand{\theequation}{S\arabic{equation}}
\renewcommand{\thefigure}{S\arabic{figure}}

\onecolumngrid

\begin{center}
  {\Large Supplemental Material for: \\ \mytitle}

\vspace*{0.5cm}
Fernando Iemini, Christophe Mora and Leonardo Mazza
\vspace{0.25cm}
\end{center}

In this Supplemental Material we provide additional information 
which have been omitted from the main text.

\section{Fock Parafermions}\label{Sec:Fock:Parafermions}

Since in the text we use a slightly different convention from that introduced in Ref.~\cite{Cobanera_2014_S}, we recap here the main formulas characterizing the relation between usual parafermions and Fock parafermions.

The easist way to express the relation between usual and Fock parafermions is the following:
\begin{equation}
 \hat \gamma_{2j-1} = \omega \left( 
 \hat C_j + \hat C_j^{\dagger2}
 \right); \qquad 
 \hat \gamma_{2j} = \hat C_j \omega^{\hat N_j} +\hat C_{j}^{\dagger2};
 \qquad 
 \hat N_j = \hat C^\dagger_j \hat C_j+\hat C^{\dagger2}_j \hat C_j^2.
\end{equation}
These equations can be inverted, yielding:
\begin{subequations}
\begin{eqnarray}
 \hat C_j &=& \frac 23 \omega^* \hat \gamma_{2j-1} - \frac 13 \hat \gamma_{2j} - \frac 13 \omega^* \hat \gamma_{2j-1}^\dagger \hat \gamma_{2j}^\dagger; \qquad
 \hat C^\dagger_j = \frac 23 \omega \, \hat \gamma_{2j-1}^\dagger - \frac 13 \hat \gamma_{2j}^\dagger - \frac 13 \omega \, \hat \gamma_{2j} \hat \gamma_{2j-1}; \\
 \hat C_j^{2} &=& \frac 13 \omega \, \hat \gamma_{2j-1}^\dagger + \frac 13 \hat \gamma_{2j}^\dagger + \frac 13 \omega \, \hat \gamma_{2j}\hat \gamma_{2j-1} ; \qquad
 \hat C_j^{\dagger2} = \frac 13 \omega^* \hat \gamma_{2j-1}+ \frac 13 \hat \gamma_{2j} + \frac 13 \omega^* \hat \gamma_{2j-1}^\dagger \hat \gamma_{2j}^\dagger.
\end{eqnarray}
\end{subequations}
From this follows that:
\begin{equation}
\hat C_j^\dagger \hat C_j = \frac 23 - \frac 13 \omega \, \hat \gamma^\dagger_{2j-1} \hat \gamma_{2j} - \frac 13 \omega^* \hat \gamma_{2j}^\dagger \hat \gamma_{2j-1};
\qquad 
\hat C_j^{\dagger2} \hat C_j^2 = \frac 13 + \frac 13 \omega^*  \hat \gamma^\dagger_{2j-1} \hat \gamma_{2j} + \frac 13 \omega \, \hat \gamma_{2j}^\dagger \hat \gamma_{2j-1};
\end{equation}
and that:
\begin{equation}
\hat N_j = 1  + \frac 13 (\omega^*-\omega) \, \hat \gamma^\dagger_{2j-1} \hat \gamma_{2j} - \frac 13 (\omega-\omega^*) \hat \gamma_{2j}^\dagger \hat \gamma_{2j-1} = 
1  - i\frac {\sqrt 3}3  \hat \gamma^\dagger_{2j-1} \hat \gamma_{2j} + i\frac {\sqrt 3}3 \hat \gamma_{2j}^\dagger \hat \gamma_{2j-1}.
\end{equation}

\section{Exactly-solvable point for $\phi = 0$}

We demonstrate that $\hat \ell_j \ket{g_{i,\phi = 0}} = 0$.
We recall the explicit expression for $\hat \ell_j = \omega^{- \hat N_j} \hat C^\dagger_j - \hat C_{j+1}^\dagger+ \hat C^2_{j}-\hat C^2_{j+1}$ and for the state: 
\begin{equation}
 \ket{g_{i, \phi = 0}} = \frac{1}{\sqrt{\mathcal N_{L,i}}}
 \sum_{\sum n_j \text{mod} 3 = i} \ket{\{ n_j\}},
 \quad i = 0,1,2.
\end{equation}
Let us consider for example $\ket{g_{0,\phi = 0}}$ (the discussion for the other states is analogous).
After the action of $\hat \ell_j$, the new state is the linear superposition of some (not all!) configurations with number of fermions $N \mod3 = 1$. 
If we focus on the sites $j$ and $j+1$, the state $\hat \ell_j \ket{g_{0,\phi = 0}}$ can be expanded in Fock states $\ket{ \{ n_j \} }$ which at those sites can have configurations with:
\begin{equation}
 \ket{\cdots \; \star \star\; \cdots} 
 \quad
 \ket{\cdots \; \star \bullet\; \cdots} 
 \quad
 \ket{\cdots \; \bullet \star\; \cdots} 
 \quad
 \ket{\cdots \; \star \circ \; \cdots} 
 \quad
 \ket{\cdots \; \bullet \bullet\; \cdots} 
 \quad
 \ket{\cdots \; \circ \star \; \cdots} 
 \quad
 \ket{\cdots \; \bullet \circ \; \cdots} 
 \quad
 \ket{\cdots \; \circ \bullet\; \cdots} 
 \quad
 \ket{\cdots \; \circ \circ \; \cdots} ;
\end{equation}
where $\star$ means ``double-occupied site'' $\bullet$ means ``single-occupied site'' and $\circ$ means ``empty site''.

We now consider a particular configuration 
$\ket{ \{ \bar j_n \} }$ which belongs to the class $\ket{\cdots \; \star \star\; \cdots}$ and demonstrate that $\hat \ell_j \ket{g_{i,\phi = 0}}$ has no overlap with such Fock state. 
The configuration $\ket{ \{ \bar j_n \} }$ may appear in $\hat \ell_j \ket{g_{i,\phi = 0}}$ as the result of the action of $\hat C^\dagger_j$ or $\hat C_{j+1}^\dagger$ on related configurations. If $j$ is even (in this and in the following examples we do not write explicitly phases which can be factored out):
\begin{equation}
+ \omega^{- \hat N_j} \hat C_j^\dagger 
\ket{\cdots \; \bullet \star\; \cdots} = 
 + \omega \, \ket{\cdots \; \star \star\; \cdots} ;
\qquad
- \hat C_{j+1}^\dagger 
\ket{\cdots \; \star \bullet \; \cdots} =
 - \omega
 \ket{\cdots \; \star \star\; \cdots} .
\end{equation}
The action of the two operators interferes destructively and thus no Fock state with two neighboring fermions at sites $j$ and $j+1$ can appear. 
Similar reasoning apply for the other states, as we list here below:
\begin{subequations}\begin{align}
+ \omega^{- \hat N_j} \hat C_j^\dagger 
\ket{\cdots \; \bullet \bullet \; \cdots} = 
+ \omega \ket{\cdots \; \star\bullet \; \cdots},
&\qquad
-\hat C^\dagger_{j+1} 
\ket{\cdots \; \star \circ \; \cdots} =
- \omega\ket{\cdots \; \star\bullet \; \cdots} ;\\ 
 + \omega^{- \hat N_j}\hat C_j^\dagger 
 \ket{\cdots \; \circ \bullet \; \cdots} = 
 + \omega^* \ket{\cdots \; \bullet \star\; \cdots} ,
 &\qquad
 -\hat C^\dagger_{j+1} 
 \ket{\cdots \; \bullet \bullet \; \cdots} =
  - \omega^* 
  \ket{\cdots \; \bullet \star \; \cdots} ;\\ 
 + \omega^{- \hat N_j}\hat C_j^\dagger 
 \ket{\cdots \; \bullet \circ \; \cdots} = 
 + \omega \ket{\cdots \; \star \circ\; \cdots} ,
 &\qquad
 -\hat C^2_{j+1} 
 \ket{\cdots \; \star \star \; \cdots} =
  - \omega
  \ket{\cdots \; \star \circ \; \cdots} ;\\ 
 + \omega^{- \hat N_j}\hat C_j^\dagger 
 \ket{\cdots \; \circ \bullet \; \cdots} = 
 + \omega^*
  \ket{\cdots \; \bullet\bullet \; \cdots} , 
  &\qquad
  -\hat C^\dagger_{j+1} 
  \ket{\cdots \; \bullet\circ \; \cdots} =
   - \omega^*
   \ket{\cdots \; \bullet\bullet \; \cdots} ;\\ 
 + \hat C_j^2 
 \ket{\cdots \; \star \star \; \cdots} = 
 + 
  \ket{\cdots \; \circ \star \; \cdots} , 
  &\qquad
  -\hat C^\dagger_{j+1} 
  \ket{\cdots \; \circ \bullet \; \cdots} =
  - 
  \ket{\cdots \; \circ \star \; \cdots} ;\\ 
 + \omega^{- \hat N_j}\hat C_j^\dagger 
 \ket{\cdots \; \circ \circ \; \cdots} = 
 + \omega^*
  \ket{\cdots \; \bullet \circ\; \cdots} , 
  &\qquad
 -\hat C^2_{j+1} 
  \ket{\cdots \; \bullet \star \; \cdots} =
  - \omega^*
  \ket{\cdots \; \bullet \circ \; \cdots} ;\\ 
 +\hat C_j^2 
 \ket{\cdots \; \star \bullet \; \cdots} = 
 + 
 \ket{\cdots \; \circ \bullet\; \cdots} , 
 &\qquad
 -\hat C^\dagger_{j+1} 
 \ket{\cdots \; \circ \circ \; \cdots} =
 - 
 \ket{\cdots \; \circ \bullet \; \cdots} ;\\ 
 + \hat C_j^2
 \ket{\cdots \; \star \circ \; \cdots} = 
 + 
 \ket{\cdots \; \circ \circ \; \cdots} , 
 &\qquad
 -\hat C^2_{j+1} 
 \ket{\cdots \; \circ  \star \; \cdots} =
 - 
 \ket{\cdots \; \circ \circ \; \cdots} .
 \end{align}
\end{subequations}

This demonstrates the property $\hat \ell_j \ket{g_{0,\phi = 0}} = 0$. The reader can easily convince himself that the same property would hold for $\ket{g_{1,\phi = 0}}$ and $\ket{g_{2,\phi = 0}}$.

The excited eigenstates of the model can be obtained by applying the $\big(\hat{\ell}_j^{\dagger} \big)^m$ ($m = 1,2$)
operators to the ground state, which introduce an energy excitation equal to  $2J\left[1-\Re(\omega^{m})\right]$.
From this follows that the gap of the model is $3J$. 
Generally speaking, the excited eigenstates and eigenvalues of the problem are given by,
\begin{equation}
\hat{H} \,\left(\hat{\ell}_1^{\dagger m_1}...\hat{\ell}_{L}^{\dagger m_{L}} |g_{i,\phi=0}\rangle\right) =
E_{m_1...m_L}  \,\left(\hat{\ell}_1^{\dagger {m_1}}...\hat{\ell}_{L}^{\dagger {m_{L}}} |g_{i,\phi=0}\rangle\right);
\qquad 
E_{m_1...m_L} = \sum_{j=1}^{L-1} 2J\left[1-\Re(\omega^{m_j})\right].
\end{equation}
Note, however, that these eigenstates are not normalized. 
 
The demonstration of the previous relation is given by recursively using the ground state identity, 
\begin{equation}
\hat \ell_j \ket{g_{i,\phi = 0}} = 0 \; \Longrightarrow \; \hat \gamma_{2j}^\dagger \ket{g_{0,\phi = 0}}
 = \omega \, \hat \gamma_{2j+1}^\dagger \ket{g_{i,\phi = 0}},
\end{equation}
or in its more general form 
\begin{equation}
(\hat \gamma_{2j}^{\dagger })^m \ket{g_{0,\phi = 0}}
 = \alpha_m \, (\hat \gamma_{2j+1}^{\dagger })^m \ket{g_{0,\phi = 0}},
 \qquad 
 \alpha_m = \omega^{ m^2  - \sum_{j=1}^{m-1} j}.
 \label{eq:gs.identity}
\end{equation}
Let us consider, for simplicity, an eigenstate with a single excitation $\big(\hat{\ell}_{ j}^\dagger \big)^{m} |g_{i,\phi=0}\rangle$, which by using Eq.~\eqref{eq:gs.identity}, could also be written in the following form,
\begin{eqnarray}
\big(\hat{\ell}_{j}^{\dagger} \big)^m |g_{i,\phi=0}\rangle &=& (\hat \gamma_{2j} - w^* \hat \gamma_{2j+1})^m |g_{i,\phi=0}\rangle =  \vartheta_m \hat \gamma_{2j}^{m} |g_{i,\phi=0}\rangle,
\end{eqnarray} 
with  $\vartheta_m $ a phase (irrelevant for our purposes) arising from Eq.~\eqref{eq:gs.identity} and  the commutation relations of
 the parafermions.
   Since $\left[\hat{\ell}_j^\dagger \hat{\ell}_j, \hat{\ell}_{k\neq j}^{\dagger^m}\right] = 0$, 
  we must only consider the Hamiltonian term at the site $j$, namely:
  \begin{eqnarray}
\left( \hat{\ell}_j^\dagger \hat{\ell}_j\right)  \big(\hat{\ell}_{ j}^{\dagger} \big)^m |g_{i,\phi=0}\rangle  = 
\left( 2 \mathbb{I} - w^* \hat \gamma_{2j+1} \hat \gamma_{2j}^\dagger -
 w \hat \gamma_{2j}\hat \gamma_{2j+1}^\dagger \right)\, \vartheta_m \hat \gamma_{2j}^m |g_{i,\phi=0}\rangle.
  \end{eqnarray}
By using again Eq.~\eqref{eq:gs.identity}, we obtain $\left( \hat{\ell}_j^\dagger \hat{\ell}_j\right)   \hat{\ell}_{ j}^{\dagger^m} |g_{i,\phi=0}\rangle =  2J\left[1-Re(w^{m})\right]\hat{\ell}_{ j}^{\dagger^m}|g_{i,\phi=0}\rangle$, which concludes the demonstration.

\section{Symmetries of the model}

In order to discuss the symmetries of the model in a transparent way, we introduce the local operators $\hat \sigma_j$ and $\hat \tau_j$ ($j$ is a site index) whose action on the local Hilbert space is represented by the following matrices:
\begin{equation}
\sigma = \left( \begin{matrix}
0 & 1 & 0 \\ 0 & 0 & 1 \\ 1 & 0 & 0
                \end{matrix} \right);
                \quad 
 \tau = \left( \begin{matrix}
1 & 0 & 0 \\ 0 & \omega & 0 \\ 0 & 0 & \omega^2                    
                    \end{matrix} \right);
                    \quad 
\omega = e^{2 \pi i / 3}.
\end{equation}
The operators satisfy the following algebra:
\begin{equation}
 \hat \sigma_j^3 = 1; \quad \hat \tau_j^3 = 1; \quad 
 \hat \sigma_j \hat \tau_j = \omega \hat \tau_j \hat \sigma_j; \quad 
 \hat \sigma_j \hat \tau_k = \hat \tau_k \hat \sigma_j \text{ for } j \neq k. 
\end{equation}
We then introduce the Fradkin-Kadanoff transformation, which is a unitary non-local transformation from parafermions to models written in terms of the operators $\{\sigma_j, \tau_j \}$:
\begin{equation}
 \hat \gamma_{2j-1} = \left(
 \prod_{k = 1}^{j-1} \hat \tau_k \right) 
 \hat \sigma_j;
 \qquad 
 \hat \gamma_{2j} = \omega \left(
 \prod_{k = 1}^{j-1} \hat \tau_k \right) 
 \hat \sigma_j \hat \tau_j.
\end{equation}
After applying it to the model $\hat H_0 + b \hat H_1 + b^2 \hat H_2$ written in the main text, we obtain the following Hamiltonian:
\begin{equation}
 \hat H = \sum_j \left[ -f \hat \tau_j  -J \omega^* \hat \upsilon_{b,j}^\dagger \hat \upsilon_{b,j+1} + \mathrm{H.c.}\right]; 
 \qquad
 \hat \upsilon_{b,j} = \hat \sigma_j^\dagger + b \left( \omega^* \, \hat \tau_j \hat \sigma_j^\dagger +  \omega \, \hat \tau_j^\dagger \hat \sigma_j^\dagger \right).
 \label{eq:Ham.full}
\end{equation}
In order to better connect with the discussions which have already presented on the symmetries of parafermionic models (see e.g.\ Ref.~\cite{Mong_2014_S}), we apply the following unitary and canonical transformation: 
$\hat \sigma_j \to \omega^{-j} \hat \sigma_j$ and $\hat \tau_j \to \hat \tau_j$ which absorbs the phase multiplying the $J$ term of the Hamiltonian.

Let us first remark that the model is invariant under the $\mathbb Z_3$ transformation $\hat Q = \prod_j \hat \tau_j^\dagger$. This follows from the fact that: (i) $\hat Q \hat \tau \hat Q^\dagger = \hat \tau$ and $\hat Q \hat \sigma \hat Q^\dagger = \omega \hat \sigma$, and (ii) the $\hat \sigma_j$ appear only combinations $\hat \sigma_j \hat \sigma_{j+1}^\dagger$.
After the transformation discussed above, the model is also explicitly invariant under spatial inversion.

The discussion of time-reversal symmetry and charge conjugation requires more care. We first introduce the following time-reversal anti-unitary transformation: $\mathcal T[\hat \tau_j] = \hat \tau_j^\dagger$, $\mathcal T[\hat \sigma_j] = \hat \sigma_j$; the $f$ part of the Hamiltonian is explicitly invariant. To demonstrate that also the $J$ part is, we observe that:
\begin{equation}
 \mathcal T [\hat v_{b,j}] = 
 \mathcal T \left[ \hat \sigma_j^\dagger + b \left( \omega^* \, \hat \tau_j \hat \sigma_j^\dagger +  \omega \, \hat \tau_j^\dagger \hat \sigma_j^\dagger \right) \right] = 
 \hat \sigma_j^\dagger + b \left( \omega \, \hat \tau_j^\dagger \hat \sigma_j^\dagger +  \omega^* \, \hat \tau_j \hat \sigma_j^\dagger \right) = \hat v_{b,j}.
\end{equation}

Charge conjugation is defined in the following way: $\mathcal C [ \hat \sigma_j ] = \hat \sigma_j^\dagger$ and $\mathcal C [ \hat \tau_j ] = \hat \tau_j^\dagger$.
Thus, as soon as $b \neq0$ the Hamiltonian is not charge-conjugation invariant. We can make sense of this by thinking at the action of $\mathcal C[\cdot]$ on the Fock-parafermion number operator defined in the main text and explicitly discussed in the Section~\ref{Sec:Fock:Parafermions} of the Supplemental Material; once the Fradkin-Kadanoff transformation is applied, we obtain:
\begin{equation}
 \hat N_j = 1 + \frac 13 \left[ (\omega^* - \omega) \hat \tau_j  + (\omega-\omega^*) \hat \tau_j^\dagger\right] = \left(  \begin{matrix}
  1 & 0 & 0 \\ 0 & 0 & 0 \\ 0 & 0 & 2
 \end{matrix} \right);
 \qquad 
 \mathcal C[\hat N_j] =  \left(  \begin{matrix}
  1 & 0 & 0 \\ 0 & 2 & 0 \\ 0 & 0 & 0
 \end{matrix} \right).
\end{equation}
Thus, charge conjugation acts as expected by swapping the states with $0$ and $2$ Fock parafermions.
Since the ground states $\ket{g_{i,\phi}}$ discussed in the main text are indeed states where the average number of Fock parafermions varies between the empty to the full situation, it does not come as a surprise that the Hamiltonian is not charge-conjugation invariant for $b \neq 0$ (which is to say $\phi \neq 0$).

As a final remark, let us mention that the composition  $\mathcal T \circ \mathcal C$ defines a time-reversal symmetry which is different from the previous one. It is anti-unitary and acts on the operators as $\mathcal T \circ \mathcal C [ \hat \sigma_j ] = \hat \sigma_j^\dagger$ and $\mathcal T \circ \mathcal C [ \hat \tau_j ] = \hat \tau_j$. From the previous considerations, it follows that for $b \neq 0$ the system is not invariant with respect to this second time-reversal symmetry.

\section{ Analytical properties of the ground state wavefunctions }

\subsection{The wavefunction}

We evaluate the properties of the ground state $\ket{g_{i,\phi}}$ in Eq.~(7) of the main text for generic value of $\phi$, which we rewrite here for reading convenience:
\begin{equation}
| g_{i,\phi} \rangle = \frac{1}{\sqrt{{\cal N}_{L,\phi,i}}} \sum_{\sum n_j \text{mod}3=i} e^{- \phi N/3} | \{ n_j\} \rangle;
\qquad i = 0,1,2.
\end{equation}
The normalization coefficient is given by the following expression
\begin{equation}
{\cal N}_{L,\phi,i} = 
\sum_{\sum n_j \text{mod} 3 = i} e^{- 2 \phi N/3} \sum_{n_1+2 n_2 = N} \binom{L}{n_1,n_2,L-n_1-n_2}.
\label{eq:norm}
\end{equation}
The multinomial coefficient is defined as:
\begin{equation}
 \binom{L}{n_1,n_2,L-n_1-n_2} = \frac{L!}{n_1! n_2! (L-n_1-n_2)!}
\end{equation}
and counts the number of ways in which $N$ Fock-parafermions can be distributed among the $L$ sites; the counting considers that there are $n_1$ sites occupied by one parafermion and $n_2$ sites occupied by two parafermions, and $N = n_1 + 2n_2$.

The expression~\eqref{eq:norm} can be simplified using the following formulas:
\begin{subequations}
\begin{align}
\left( 1 + e^{-2 \phi/3} +  e^{-4 \phi/3} \right)^L & = \sum_{n_1,n_2}  \binom{L}{n_1,n_2,L-n_1-n_2} \left( e^{-2 \phi/3} \right)^{n_1 + 2 n_2}; \\
\left( 1 + \omega e^{-2 \phi/3} + \omega^2 e^{-4 \phi/3} \right)^L & = \sum_{n_1,n_2}  \binom{L}{n_1,n_2,L-n_1-n_2} \left( \omega \, e^{-2 \phi/3} \right)^{n_1 + 2 n_2};\\
\left( 1 + \omega^2 e^{-2 \phi/3} + \omega^4 e^{-4 \phi/3} \right)^L & = \sum_{n_1,n_2}  \binom{L}{n_1,n_2,L-n_1-n_2} \left( \omega^2 e^{-2 \phi/3} \right)^{n_1 + 2 n_2};\\
1+ \omega^{n_1+2 n_2} + (\omega^2)^{n_1+2 n_2} & = 3 \, \delta_{(n_1 + 2 n_2) \text{mod}3 = 0}.
\end{align}
\label{eq:complicatedformulas}
\end{subequations}
We obtain the following result:
\begin{equation} 
{\cal N}_{L,\phi,0} = \frac 13 \sum_{k=0}^{2}  A (L,\phi,k), \qquad 
{\cal N}_{L,\phi,1} = \frac 13 \sum_{k=0}^{2} \omega^{-k} A (L,\phi,k), \qquad
{\cal N}_{L,\phi,2} = \frac 13 \sum_{k=0}^{2} \omega^{-2 k} A (L,\phi,k),
\label{eq:normalizzazioni}
\end{equation}
where we have introduced the notation $A(L,\phi,k) = \left( 1+ \omega^k e^{-2\phi/3} + \omega^{2 k} e^{-4\phi/3} \right)^L$. At $\phi = 0$, these expressions reduce to ${\cal N}_{L,\phi,0} = {\cal N}_{L,\phi,1} = {\cal N}_{L,\phi,1} = 3^{L-1}$.

\subsection{Expectation value of a $\mathbb Z_3$-preserving observable}

Using this explicit form of the ground state wavefunction, we can compute the average density on a given site $j$. The number operator for site $j$ is given by $\hat{N}_j = C^\dagger_j C_j + C^{\dagger}_j C^{\dagger}_j C_j C_j$. As an example of how the calculation works, let us compute the probability $p_{0,\phi}(1_j)$ for the state $\ket{g_{0,\phi}}$ that there is one and only one parafermion at site $j$:
\begin{equation} 
p_{0,\phi}(1_j) = \frac{1}{{\cal N}_{L,\phi,0}} 
\sum_{\sum n_j \text{mod}3 = 2}
\sum_{n_1+2 n_2 = N - 1  } \binom{L-1}{n_1,n_2,L-1-n_1-n_2} e^{-2 \phi N/3} =  \frac{e^{-2\phi/3}}{3 \, {\cal N}_{L,\phi,0} } \sum_{k=0}^2 \omega^{-2 k} A(L-1,\phi,k).
\end{equation}
The calculation of $\langle \hat N_j \rangle$ follows these lines and we obtain
\begin{align}
\langle g_{0,\phi} | \hat{N}_j  | g_{0,\phi} \rangle & = \frac{e^{-2\phi/3} \sum_{k=0}^2 \omega^{-2 k} A(L-1,\phi,k) + 2 e^{-4\phi/3} \sum_{k=0}^2 \omega^{-k} A(L-1,\phi,k)}{\sum_{k=0}^2  A(L,\phi,k)} \\
\langle g_{1,\phi} | \hat{N}_j  | g_{1,\phi} \rangle & = \frac{e^{-2\phi/3} \sum_{k=0}^2  A(L-1,\phi,k) + 2 e^{-4\phi/3} \sum_{k=0}^2 \omega^{-2 k} A(L-1,\phi,k)}{\sum_{k=0}^2 \omega^{-k} A(L,\phi,k)} \\
\langle g_{2,\phi} | \hat{N}_j | g_{2,\phi} \rangle & = \frac{e^{-2\phi/3} \sum_{k=0}^2  \omega^{-k} A(L-1,\phi,k) + 2 e^{-4\phi/3} \sum_{k=0}^2 A(L-1,\phi,k)}{\sum_{k=0}^2 \omega^{-2 k} A(L,\phi,k)}
\end{align}
For large system size, $L \gg 1$, the mean densities are all equal 
\begin{equation}
\langle g_{i,\phi} | \hat{N}_j  | g_{i,\phi} \rangle \simeq \frac{e^{-2\phi/3}+2 e^{-4\phi/3}}{1+e^{-2\phi/3} + e^{-4\phi/3}},
\end{equation}
corresponding to $\langle  \hat{N}_j   \rangle = 1$ for $\phi = 0$. Noting that $A(L,\phi,1) = A^*(L,\phi,2)$ ($A(L,\phi,0)$ is real), the large $L$ asymptotics can be obtained from
\begin{equation}
\left| \frac{A(L,\phi,1)}{A(L,\phi,0)} \right| = e^{-L/\xi}
\label{eq:rapporto}
\end{equation}
with the length
\begin{equation}
\xi^{-1} = \ln \left| \frac{1+e^{-2\phi/3} + e^{-4\phi/3}}{ 1+ \omega e^{-2\phi/3} + \omega^* e^{-4\phi/3}} \right|.
\end{equation}
The first order correction to the mean densities then takes the form
\begin{equation}
\langle g_{i,\phi} | \hat{N}_j  | g_{i,\phi} \rangle = \frac{e^{-2\phi/3}+2 e^{-4\phi/3}}{1+e^{-2\phi/3} + e^{-4\phi/3}} + \frac{e^{-2\phi/3}}{1+e^{-2\phi/3} + e^{-4\phi/3}} \, {\cal O} \left( e^{-L/\xi} \right) + \frac{e^{-4\phi/3}}{1+e^{-2\phi/3} + e^{-4\phi/3}}  \,  {\cal O} \left( e^{-L/\xi} \right).
\end{equation}
The infinite length result is thus approached with exponential accuracy as expected for a topologically protected manifold. Interestingly, the characteristic length $\xi$ diverges for $\phi \to \pm\infty$ corresponding to a loss of topological protection.

\subsection{Correlation function}

Let us now use the knowledge on the ground state wavefunction to compute correlation functions. As we shall see, they involve the same correlation length $\xi$. To be specific, we concentrate on the following function
\begin{equation}
G_0(\ell) = \langle g_{0,\phi} | \hat{C}_1^{\dagger \, 2} \hat{C}_\ell^2 | g_{0,\phi} \rangle
\end{equation}
measuring the two-particle correlation between site $1$ (left boundary) and $\ell$. $\hat{C}_j$ denotes a Fock-parafermion operator at site $1 \le j \le L$. To be non-vanishing, the part of $| g_{0,\phi} \rangle$ onto which $\hat{C}_\ell^2$ acts must contain two particles at site $\ell$ whereas the bra $\langle g_{0,\phi} |$ must contain two {zero} particles at site $1$. Taking into account the commutator, 
\begin{equation}
\hat{C}_\ell^2 \hat{C}_j^\dagger = \omega^2 \hat{C}_j^\dagger \hat{C}_\ell^2 = \omega^* \hat{C}_j^\dagger \hat{C}_\ell^2 \qquad \qquad \text{for} \quad \ell >j,
\end{equation}
containing information on particle statistics, we obtain 
\begin{equation}
G_0(\ell) = \frac{1}{{\cal N}_0} \sum_{n_1+2 n_2 = N_1 \atop n_3 + 2 n_4 = N_2 \quad N_1 + N_2 + 2 = 0[3]}  
\left(\omega^* \right)^{N_1} \binom{\ell -2}{n_1,n_2,\ell-2-n_1-n_2} \binom{L-\ell}{n_3,n_4,L-\ell-n_3-n_4}  e^{- \frac{2 \phi}{3} (N_1+N_2+2)}.
\end{equation}
 In this expression, $N_1$ is the number of particles enclosed between site $1$ and site $\ell$, and $N_2$ the number of particles beyond site $\ell$. The total number of particles is therefore $N_1+N_2+2$. The $\left(\omega^* \right)^{N_1}$ term is the statistical phase steming from the commutation of $\hat{C}_\ell^2$ with $N_1$ particles. Instead of enforcing the constraint $N_1+N_2+2\equiv 0 \; \mod 3$, we use the same trick as above, with the summation
\begin{equation}
\frac{1}{3} \left( 1+ \omega^N + (\omega^2)^N \right) = \delta_{N (\text{mod} 3) , 0} \; ,
\end{equation}
to express the correlation function as a sum over a new index $k$,
\begin{equation}
G_0(\ell) = \frac{1}{{\cal N}_0} \sum_{k=0}^2 \sum_{n_1+2 n_2 = N_1 \atop n_3 + 2 n_4 = N_2}  \omega^{k(N_1+N_2+2)}
\left(\omega^* \right)^{N_1} \binom{\ell -2}{n_1,n_2,\ell-2-n_1-n_2} \binom{L-\ell}{n_3,n_4,L-\ell-n_3-n_4}  e^{- \frac{2 \phi}{3} (N_1+N_2+2)}.
\end{equation}
With this new formulation, the summations over $N_1$ and $N_2$ are decoupled and can be performed exactly using the binomial formulas~\eqref{eq:complicatedformulas}. The result takes the form
\begin{equation}
G_0(\ell) = e^{-4 \phi/3} \frac{\sum_{k=0}^2 \omega^{2 k} A(\ell-2,\phi,k-1) A(L-\ell,\phi,k)}{\sum_{k=0}^2  A(L,\phi,k)}
\end{equation}
with again the notation 
$A(L,\phi,k) = \left( 1+ w^k e^{-2 \phi/3} + w^{2 k} e^{-4 \phi/3} \right)^L$.

Let us take a few examples. 
\begin{enumerate}
\item Short range $\ell =1$ in the large length limit $L\gg 1$.
\begin{equation}
G_0(1) =  \langle g_0 |  \hat{C}_1^{\dagger} \hat{C}_1^{\dagger}  \hat{C}_1 \hat{C}_1 | g_0 \rangle = \frac{e^{-4 \phi/3}}{1+e^{-2 \phi/3}+e^{-4 \phi/3}} \to \frac{1}{3} \qquad \textrm{for} \quad \phi \to 0
\end{equation}
It measures the probability to find two particles at site $1$ in the ground state. For $\phi=0$, the three occupations zero, one and two are equiprobable.
\item Next neighbour $\ell =2$ in the large length limit $L\gg 1$.
\begin{equation}
G_0(2) =  \langle g_0 |  \hat{C}_1^{\dagger} \hat{C}_1^{\dagger}  \hat{C}_2 \hat{C}_2 | g_0 \rangle = \frac{e^{-4 \phi/3}}{\left(1+e^{-2 \phi/3}+e^{-4 \phi/3}\right)^2} \to \frac{1}{9} \qquad \textrm{for} \quad \phi \to 0
\end{equation}
$G_0(2)$ gives the probability to find zero particle at site $1$ and two particles at site $2$ (or the opposite), so $1/9$ when all occupations are equiprobable.
\item End site $\ell = L$ in the large length limit $L\gg 1$. The result is 
\begin{equation}
G_0(L) = \langle g_0 |  \hat{C}_1^{\dagger} \hat{C}_1^{\dagger}  \hat{C}_L \hat{C}_L | g_0 \rangle = \frac{\omega^2 \, e^{-4 \phi/3}}{\left(1+e^{-2 \phi/3}+e^{-4 \phi/3}\right)^2} = \omega^2 \, G_0(2)
\end{equation}
The correlation over the whole system is finite (a signature of non-locality related to topology), and equal to $G_0(2)$ up to a phase shift $\omega^2$. This phase shift depends in fact on the parity of the ground state and one finds $G_1(L)/G_1(2) = \omega$ and $G_2(L)/G_2(2) = 1$.
\item Asymptotic decay. We first consider the limit $L \to \infty$ and then take the limit $\ell \gg 1$ to avoid short-range details. One finds
\begin{equation}
| G_0(\ell) | = \frac{e^{-4 \phi/3} \, e^{2/\xi}}{\left(1+e^{-2 \phi/3}+e^{-4 \phi/3}\right)^2} \, e^{-\ell/\xi} \sim e^{-\ell/\xi} \qquad \textrm{for} \quad \ell \gg 1,
\end{equation}
{\it i.e.} an exponential decay with the correlation length $\xi$. This result does not depend on the parity and applies also when the correlation function is averaged over $G_1(\ell)$ and $G_2(\ell)$. A difference occurs in the phase of but not in the modulus.
\end{enumerate}

\subsection{Expectation value of a $\mathbb Z_3$-breaking observable}

Let us now compute the expectation value of a $\mathbb{Z}_3$-breaking observable. To be specific, we concentrate on the following function,
\begin{eqnarray}
F_{i}(\ell) &=& \langle g_{i,\phi}| \hat C_\ell^\dagger |g_{i',\phi}\rangle;
\qquad i' \equiv i-1 (\text{mod} 3)
\end{eqnarray}
In order to be different from zero, the Fock states upon which $|g_{i',\phi}\rangle$ is expanded which are relevant for the calculation must have zero (one) particle at   site $\ell$, whereas for $\langle g_{i,\phi}|$ they must contain one (two) particle at the same site $\ell$. 
Taking into account the commutator,
 \begin{equation}
 \hat C_\ell^\dagger \hat C_j^\dagger = \omega^* \hat C_j^\dagger \hat C_\ell^\dagger  \quad \mbox{for } \ell > j,
 \end{equation}
  we obtain
 \begin{equation}
F_{i}(\ell) =
  \frac{1}{\sqrt{\mathcal{N}_{L,\phi,i} \, \mathcal{N}_{L,\phi,i'} } } \hspace{-0.7cm}
\sum_{ \scriptsize{\begin{array}{cccc} n_\ell = 0,1\\ n_1+2n_2 = N_1\\ n_3+2n_4=N_2\\ N =N_1+N_2+n_\ell\equiv i-1 \,(\text{mod}3)\\  \end{array}} } \hspace{-0.7cm}
(\omega^*)^{N_1} \binom{\ell-1}{n_1,n_2,\ell-1-n_1-n_2}\binom{L-\ell}{n_3,n_4,L-\ell-n_3-n_4}
e^{-\frac{\phi (2N+1)}{3}}. \nonumber
 \end{equation}
In this expression, $n_\ell$ is the number of particles at the site $\ell$, $N_1$ is the number of particles enclosed between site $1$ and $\ell-1$, and $N_2$ the number of particles beyond site $\ell$. 
The total number of particles in $|g_{i',\phi}\rangle$ is therefore $N = N_1+N_2+n_\ell$, whereas in $\langle g_{i,\phi}|$ must have $N+1$ particles, which yields the $e^{-\frac{\phi (2N+1)}{3}}$ term.
The $\left(\omega^*\right)^{N_1}$ term is the statistical phase stemming from the commutation of $C_\ell^\dagger$ with $N_1$ particles.
Instead of enforcing the constraint $N_1+N_2+n_\ell \equiv i-1 (\text{mod} 3)$, we use again the previous trick:
\begin{eqnarray}
\delta_{N_1+N_2+n_\ell, i-1 \,(\text{mod} 3)} = 
\frac{1}{3} \sum_{k=0}^2 \omega^{k(N_1+N_2+n_\ell - (i-1))}
\end{eqnarray}
in order to express the function as a sum over a new index $k$:
\begin{eqnarray}
F_{i}(\ell)  = 
\frac{e^{-\frac{\phi}{3} }}{3 \sqrt{\mathcal{N}_{L,\phi,i} \mathcal{N}_{L,\phi,i'} } } 
\sum_{n_\ell=0,1} \sum_{k=0}^2 \sum_{ \scriptsize{ \begin{array}{cc} n_1+2n_2 =N_1\\ n_3+2n_4 = N_2 \end{array} } }
\binom{\ell-1}{n_1,n_2,\ell-1-n_1-n_2}\binom{L-\ell}{n_3,n_4,L-\ell-n_3-n_4} \times \nonumber \\
\times \omega^{k(N_1+N_2+n_\ell - (i-1))}
(\omega^*)^{N_1}
e^{-\frac{2\phi}{3} (N_1+N_2+n_\ell) }.
\end{eqnarray}
With this new formulation, the summation over $N_1$ and $N_2$ are decoupled and can be performed 
exactly using the binomial formulas. 
The result takes the form
\begin{equation}
F_{i}(\ell) = \frac{\sum_{n_\ell=0,1} e^{-\frac{\phi}{3}(1+2 n_\ell)} \sum_{k=0}^2 \omega^{k(n_\ell - (p-1))} 
A(\ell-1,\phi,k-1) A(L-\ell,\phi,k)}{\sqrt{\sum_{k,k'=0}^2 A(L,\phi,k)A(L,\phi,k')\omega^{ik + (i-1)k'}}} 
\end{equation}

Let us analyze now its asymptotic decay. We first define the phase $\theta_{L,\phi}$ between the terms $A(L,\phi,k)$, in the large asymptotic $L$, as follows,
\begin{eqnarray}
\frac{A(L,\phi,1)}{A(L,\phi,0)} = e^{-\frac{L}{\xi}}e ^{i\theta_{L,\phi}},\qquad 
\frac{A(L,\phi,2)}{A(L,\phi,0)} = \left(\frac{A(L,\phi,1)}{A(L,\phi,0)}\right)^*
\end{eqnarray}

Two different limits are 
particularly important: 
\begin{enumerate}
\item $L \rightarrow \infty$, considering $\ell \gg \xi$, 
which allows us to analyze the ``left-side'' of 
the wire, where one finds,
\begin{eqnarray} \label{eq:Z3breaking.left}
F_{i}(\ell) &=&  (e^{\frac{\phi}{3}}+e^{-\frac{\phi}{3}}) \left[ \frac{e^{-\frac{2\phi}{3}} e^{-i\theta_{\ell-1,\phi}}  }{(1+e^{-\frac{2\phi}{3}}+e^{-\frac{4\phi}{3}})}\right] \,e^{-\frac{(\ell-1)}{\xi}} \sim e^{-\frac{\ell}{\xi}} .
\end{eqnarray}
\item $L,\ell \rightarrow \infty$, keeping $\ell/L = cte < 1-\xi$, which allows us 
to analyze the ``right-side'' of the wire, where we find similar results,
\begin{eqnarray}\label{eq:Z3breaking.right}
F_{i}(\ell) &=&  w^{*^{p}} (we^{\frac{\phi}{3}}+e^{-\frac{\phi}{3}}w^*) \, \left[ \frac{ e^{-\frac{2\phi}{3}} e^{-i\theta_{\ell-1,\phi}}  }{(1+e^{-\frac{2\phi}{3}}+e^{-\frac{4\phi}{3}})}\right]^* \,e^{-\frac{(L-\ell)}{\xi}}  \sim e^{-\frac{(L-\ell)}{\xi}} .
\end{eqnarray}
\end{enumerate}



\subsection{Entanglement spectrum}

In order to compute the entanglement spectrum of the ground states $\ket{g_{i,\phi}}$, we exploit the following property:
\begin{equation}
\ket{g_{i,\phi}} = 
\frac{1}{\sqrt{\mathcal N_{L,\phi,i}}}
\sum_{p=0}^2
\sqrt{\mathcal N_{\ell,\phi,i}}
\sqrt{\mathcal N_{L-\ell,\phi,i}}
\ket{g_{\ell,\phi,p}}
\ket{g_{L-\ell,\phi,(i-p)\text{mod}3}};
\quad i = 0,1,2.
\end{equation}
Thus, computing the entanglement spectrum for a bipartition $\ell,L-\ell$, requires the evaluation of the three coefficients:
\begin{equation}
 \frac{\mathcal N_{\ell,\phi,i} \mathcal N_{L-\ell,\phi,i}}{\mathcal N_{L,\phi,i}}.
 \label{eq:coefficients}
\end{equation}
Let us consider the case in which $\ell, L-\ell \gg \xi$. According to the expressions in Eqs.~\eqref{eq:normalizzazioni} and~\eqref{eq:rapporto}, we can simplify:
\begin{equation}
 \mathcal N_{m,\phi,i} = \frac 13 A(m,\phi,0) + \mathcal O (e^{-m/\xi});
 \quad m = L, \ell, L-\ell;
\end{equation}
where only the term scaling as $e^{-m/\xi}$ depends on $i$. At the leading order the coefficients in \eqref{eq:coefficients} do not depend on $i$ and are equal to $1/3$.
With this, we automatically obtain that the states $\ket{g_{L,\phi,i}}$ satisfy an area-law.

\subsection{Weak edge mode}

In the simple case $f=b=0$ (or $\phi=0$), the parafermionic chain has strong edge modes resulting in a global threefold degeneracy in the spectrum. The edge modes are given by the two parafermionic operators $\hat{\gamma}_1$ and $\hat{\gamma}_{2 L}$ localized at both ends of the chain. In the general case $\phi \ne0$, the edge modes are obtained by quasi-adiabatic continuation~\cite{Alexandradinata_2015_S},
\begin{equation}
\hat{\chi}_1 = \hat{\mathcal{V}}_\phi \hat{\gamma}_1 \hat{\mathcal{V}}_\phi^\dagger, \qquad \hat{\chi}_2 =  \hat{\mathcal{V}}_\phi \hat{\gamma}_{2 L} \hat{\mathcal{V}}_\phi^\dagger,
\end{equation}
with the unitary transformation $\hat{\mathcal{V}}_\phi$ mapping the low-energy subspaces at $\phi \ne 0$ and $\phi=0$. The operator $\hat{\mathcal{V}}_\phi$ is unitary and thus preserves the parafermionic properties $\hat{\chi}_1^3 =1$ and $\hat{\chi}_1^2 = \hat{\chi}_1^\dagger$, it also preserves locality such that $\hat{\chi}_1$ and $\hat{\chi}_2$ are still localized near the two ends of the chain.  $\hat{\mathcal{V}}_\phi$ furthermore commutes with the parity operator, hence interpolating between ground states of constant parity.

For $\phi \ll 1$, corresponding to $b \ll 1$ and $f \ll J$, the deformation of the edge mode can be calculated perturbatively. For the left mode, it takes the form
\begin{equation}\label{leftstate}
\hat{\chi}_1 = \hat{\gamma}_1 + \int_{-\infty}^{\infty} dt F(3 t) e^{i \hat H_0^0 t} [ \hat{V}, \hat \gamma_1] e^{-i \hat H_0^0 t},
\end{equation}
where we introduced the reduced Hamiltonian decomposition $\hat H_0^0 = \hat{H}_0 (f=0)/J$ and $\hat{V} = ( \hat{H}_0  + b \hat{H}_1)/J-\hat H_0^0$, such that $\hat{V} = 0$ at $\phi=0$. Incidentally, we have used the commutation $[\hat H_0^0,\hat \gamma_1]=0$ in deriving Eq.~\eqref{leftstate}. The filter function $F(t)$ is constraint by its Fourier transform for $|\Omega| >1$,
\begin{equation}
\tilde{F} (\Omega) = \int_{-\infty}^{\infty} \mathrm{d} t e^{i \Omega t} F(t) = -\frac{1}{\Omega}.
\end{equation}
The form of $\tilde{F} (\Omega)$ for $|\Omega| <1$ is not constraint (except that $F(t)$ must decay fast at large $t$ to preserve locality), but we will not need it to leading order in $\phi$.
 
Computing the commutator
\begin{equation}
[ \hat{V},\hat \gamma_1]  = (\omega^* - \omega) \left[ \frac{f}{J} ( \hat \gamma_2 - \omega \hat \gamma_1^\dagger \hat \gamma_2^\dagger) + b \, \omega^* (\hat \gamma_3 - \hat \gamma_1^\dagger \hat \gamma_3^\dagger ) + b \, (\omega \hat \gamma_2^\dagger \hat \gamma_3^\dagger - \hat \gamma_1^\dagger \hat \gamma_3 \hat \gamma_2) \right],
\end{equation}
we find the expansion
\begin{equation}
\hat{\chi}_1 = \hat{\gamma}_1 +  (\omega^* - \omega) \left[ \frac{f}{J} ( \hat X_2 + \omega \gamma_1^\dagger \hat X_2^\dagger) + b \, \omega^* (\hat X_3 + \hat \gamma_1^\dagger \hat X_3^\dagger ) + b \, (\omega \hat X_{23} + \hat \gamma_1^\dagger \hat X_{23}^\dagger) \right],
\end{equation}
where we have defined
\begin{subequations}
\begin{align}
\hat X_j &= \int_{-\infty}^{\infty} \mathrm{d}t F(3 t) e^{i \hat H_0^0 t} \hat \gamma_j  e^{-i \hat H_0^0 t}, \qquad \hat X_j^\dagger = - \int_{-\infty}^{\infty} \mathrm{d}t F(3 t) e^{i \hat H_0^0 t} \hat \gamma_j^\dagger  e^{-i \hat H_0^0 t}, \\
\hat X_{23} & = \int_{-\infty}^{\infty} \mathrm{d}t F(3 t) e^{i \hat H_0^0 t} \hat \gamma_2^\dagger \hat \gamma_3^\dagger  e^{-i \hat H_0^0 t},
\end{align}
\end{subequations}
and used again that $\gamma_1$ commutes with $H_0^0$ and that $F(t)$ is an imaginary function.

As discussed in Ref.~\cite{Alexandradinata_2015_S}, the time evolution of these operators can be determined from the closed algebra under commutation by $\hat H_0^0$, namely
\begin{equation}
[ \hat H_0^0, A_0 \hat \gamma_2 + B_0 \omega \hat \gamma_3 + C_0 \hat \gamma_2^\dagger \hat \gamma_3^\dagger ] =  A_1 \hat \gamma_2 + B_1 \omega \hat \gamma_3 + C_1 \hat \gamma_2^\dagger \hat \gamma_3^\dagger
\end{equation}
where the coefficients are related by
\begin{equation}
\begin{pmatrix} A_1 \\ B_1 \\ C_1 
\end{pmatrix} = 3 {\cal M} \begin{pmatrix} A_0 \\ B_0 \\ C_0 
\end{pmatrix}, \qquad \qquad {\cal M} = \frac{1}{i \sqrt{3}} \begin{pmatrix} 0 & -1 & 1 \\ 1 & 0 & -1 \\ -1 & 1 & 0 
\end{pmatrix}.
\end{equation}
The operators $\hat v_j = (\hat \gamma_2 + \omega^j \, \omega \, \hat \gamma_3 + \hat \gamma_2^\dagger \hat \gamma_3^\dagger)/\sqrt{3}$, related to the eigenstates of the matrix ${\cal M}$, follow a free evolution under $\hat H_0^0$ written as  $e^{i \hat H_0^0 t} \hat v_j e^{-i \hat H_0^0 t} = \hat v_j e^{3 i \varepsilon_i t}$, with the eigenvalues $\varepsilon_0= 0$, $\varepsilon_{1/2} = \mp 1$.

Applying the Fourier transform and the result $\tilde{F}(0)=0$ ( $\tilde{F}$ is an odd function), we are finally led to
\begin{equation}
\hat X_2 = \frac{1}{3 (\omega^* - \omega)} ( \omega \hat \gamma_3 - \hat \gamma_2^\dagger \hat \gamma_3^\dagger), \qquad \hat X_3 = \frac{\omega^*}{3 (\omega^* - \omega)} ( \hat \gamma_2^\dagger \hat \gamma_3^\dagger - \hat \gamma_2), \qquad \hat X_{23} = \frac{1}{3 (\omega^* - \omega)} ( \hat \gamma_2 - \omega \hat \gamma_3 ),
\end{equation}
and the weak edge mode takes the form
\begin{equation}\label{final}
\hat{\chi}_1 = \hat{\gamma}_1 +  \alpha ( \omega \hat \gamma_3 - \hat \gamma_2^\dagger \hat \gamma_3^\dagger) + \alpha^* (\omega \hat \gamma_1^\dagger  \hat \gamma_3 \hat \gamma_2 - \hat \gamma_1^\dagger \hat \gamma_3^\dagger )
\end{equation}
to leading order in $f/J$ and $b$. We have introduced the complex number $\alpha = f/J - \omega b$. From this result, it can be verified explicitly that $\hat{\chi}_1^2 = \hat{\chi}_1^\dagger$ and $\hat{\chi}_1^3 = 1$ to first order in $\alpha$.
For $b=0$ but $f \ne 0$, the result~\eqref{final} is similar to Ref.~\cite{Alexandradinata_2015_S}.

\end{document}